\documentclass{jps-cp}
\usepackage{txfonts,bm,braket,graphicx,xcolor}

\newcommand{\Co}{{Co$_4$Nb$_2$O$_9$}}

\newcommand{\bS}{{\bm{S}}}
\newcommand{\bp}{{\bm{p}}}
\newcommand{\bP}{{\bm{P}}}
\newcommand{\bW}{{\bm{W}}}
\newcommand{\bH}{{\bm{H}}}
\newcommand{\rA}{{\rm{A}}}
\newcommand{\rB}{{\rm{B}}}

\title{Theoretical Study of Magnetoelectric Effects \\
in Honeycomb Antiferromagnet \Co}

\author{Masashige \textsc{Matsumoto}$^{1}$ and Mikito \textsc{Koga}$^{2}$}

\inst{$^{1}$Department of Physics, Faculty of Science, Shizuoka University, Shizuoka 422-8529, Japan \\
$^{2}$Department of Physics, Faculty of Education, Shizuoka University, Shizuoka 422-8529, Japan}

\email{matsumoto.masashige@shizuoka.ac.jp}

\recdate{September 15, 2019}

\abst{
The honeycomb antiferromagnet \Co~is known to exhibit an interesting magnetoelectric effect
that the electric polarization rotates at the twice speed
in the opposite direction relative to the rotation of the external magnetic field applied in the basal $ab$-plane.
The spin-dependent electric dipole can be an origin of the magnetoelectric effect.
It is described by the product of spin operators at different sites (type-I theory) or at the same site (type-II theory).
We examine the electric polarization for the two cases on the basis of the symmetry analysis of the crystal structure of \Co,
and conclude that the latter is the origin of the observed result.
This paper also gives a general description of the field-induced electric polarization on honeycomb lattices with the $C_3$ point group symmetry
on the basis of the type-I theory.
}

\kword{ spin-dependent electric dipole, quadrupole, honeycomb lattice, $C_3$ point group symmetry}

\begin{document}
\maketitle

\section{Introduction}

\begin{figure}[b]
\begin{center}
\includegraphics[width=14cm,clip]{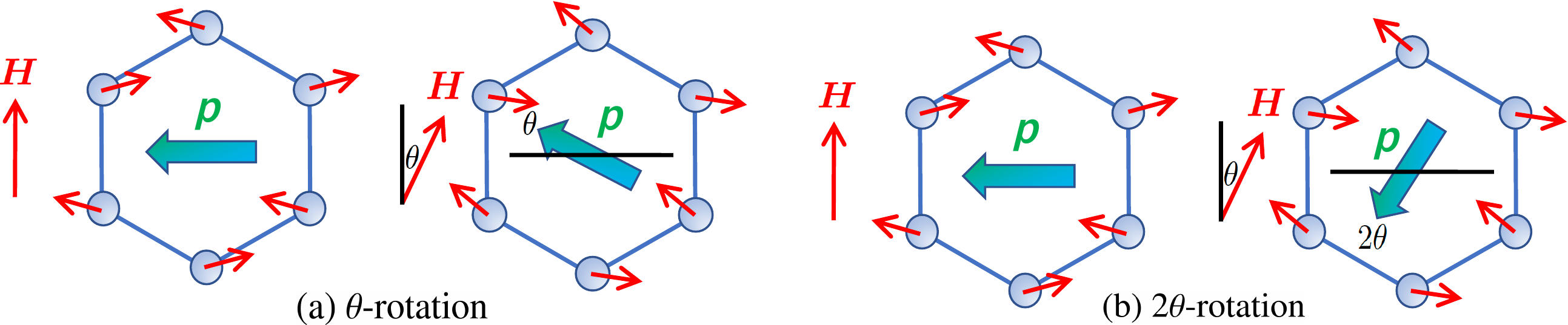}
\caption{
(Color online)
Schematic of (a) $\theta$-rotation and (b) $2\theta$-rotation of the electric polarization.
$\theta$ is the direction of the magnetic field $\bH$.
Short red arrows represent alignment of spins.
$\bP$ is the induced electric polarization.
It polarization rotates $\theta$ and $-2\theta$ under the rotation of the field for (a) and (b), respectively.
}
\label{fig:rotation}
\end{center}
\end{figure}

\Co~is known as one of typical multiferroic quantum spin systems
\cite{Fischer-1972},
where the magnetic structure is almost collinear in the basal $ab$-plane below the N\'{e}el temperature
\cite{Khanh-2016,Deng-2018}.
Under an external magnetic field applied in the $ab$-plane, it was reported that the electric polarization rotates
in the opposite direction at the twice speed relative to the rotation of the magnetic field ($2\theta$-rotation) (see Fig. \ref{fig:rotation})
\cite{Khanh-2016,Khanh-2017}.
The polarization was also reported to change its sign when the magnetic field is reversed (field-sweeping process)
\cite{Khanh-2016,Khanh-2017}.
These points were studied and explained by the first principal calculation \cite{Solovyev-2016}
and by the itinerant band picture \cite{Yanagi-2018-1,Yanagi-2018-2}.
It was also studied from a picture of quantum spin system
\cite{Matsumoto-2019}.

In general, there are two main models suggested for an origin of magnetoelectric effects,
where the electric dipole is described by product of spin operators.
The electric dipole is described by the symmetric or antisymmetric product of spin operators at different sites (Type-I theory)
\cite{Moriya-1968,Katsura-2005}.
It is also described by the product of spin operators at the same site (type-II theory)
\cite{Moriya-1968,Arima-2007}.
Symmetry analysis is one of ways to investigate the magnetoelectric effects.
It demonstrates power for complicated crystal structures such as \Co.
For the type-I theory, possible spin dependences in the electric dipole were classified by the space inversion, n-fold axis, and mirror symmetries
\cite{Kaplan-2011,Matsumoto-2017}.
For the type-II theory, it was classified by the point group symmetry
\cite{Mims-1976,Matsumoto-2017}.
In Ref. \ref{ref:Matsumoto-2019}, we reported that the observed magnetoelectric effect of the $2\theta$-rotation and the field-sweeping process
can be well explained by the type-II theory.
In this paper, we also explore another possibility on the basis of the type-I theory.

\section{Field-Induced Electric Polarization}

\begin{figure}[t]
\begin{center}
\includegraphics[width=15cm,clip]{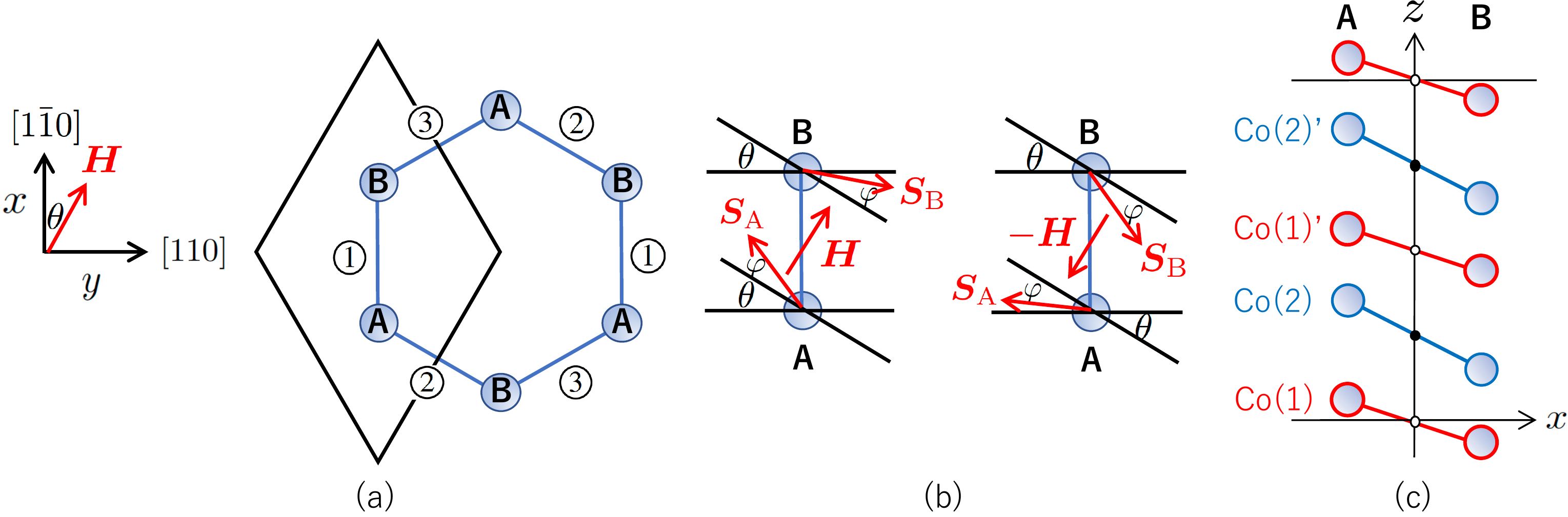}
\caption{
(Color online)
Schematic of crystal structure of \Co.
(a) Honeycomb lattice structure.
The rhombus represents the unitcell.
$x$ and $y$ axes are taken along the $[1\bar{1}0]$ and $[110]$ directions, respectively.
A and B represent the two sublattices for the up and down staggered magnetic structure, respectively.
On the honeycomb lattice, there are three kinds of bonds, as denoted by
$\textcircled{\scriptsize 1}$, $\textcircled{\scriptsize 2}$, and $\textcircled{\scriptsize 3}$.
(b) Magnetic structure in the $xy$-plane under the field $\bH$ applied in the $\theta$ direction shown in (a).
$\varphi$ represents canting angle of the spin.
When the field is reversed as $\bH\rightarrow -\bH$, it changes as $\varphi\rightarrow -\varphi$ (field-sweeping process).
(c) View of the unitcell in the $xz$-plane at the bond-$\textcircled{\scriptsize 1}$.
The open and filled circles represent the inversion center and twofold axis around the $y$ direction, respectively.
There are four honeycomb lattices in the unitcell along the $z$ direction, as denoted by Co(1), Co(1)', Co(2), and Co(2)'.
}
\label{fig:crystal}
\end{center}
\end{figure}

In the antiferromagnetic (AF) phase, the magnetic structure under an external magnetic field is shown in Fig. \ref{fig:crystal},
where the field is applied in the $xy$-plane with a finite $z$ component.
As shown in Fig. \ref{fig:crystal}(b), the expectation values of the spin operators on the two sublattices are expressed as
\cite{Matsumoto-2019}
\begin{align}
&\Bigl( \braket{S_\rA^x},\braket{S_\rA^y},\braket{S_\rA^z} \Bigr)
=
\Bigl( S_\perp \sin(\theta +\varphi),~~- S_\perp \cos(\theta +\varphi),~~S_\parallel \Bigr), \cr
&\Bigl( \braket{S_\rB^x},\braket{S_\rB^y},\braket{S_\rB^z} \Bigr)
=
\Bigl( -S_\perp \sin(\theta - \varphi),~~S_\perp \cos(\theta - \varphi),~~S_\parallel \Bigr).
\label{eqn:S}
\end{align}
Here, $S_\perp$ and $S_\parallel$ represent the amplitudes in the $xy$-plane and in the $z$ directions, respectively.
$\theta$ represents the direction of the external magnetic field, whereas $\varphi$ represents the canting angle [see Fig. \ref{fig:crystal}(b)].
There are two types of Co sites.
One has inversion center [Co(1) and Co(1)'], and the other has twofold axis around the $y$ direction [Co(2) and Co(2)'].
In the presence of the inversion center, only antisymmetric spin dependence is allowed for the electric dipole.
In the absence of the inversion center, symmetric spin dependence is also allowed in addition to the antisymmetric one.
We separately study the two cases below.

\subsection{Antisymmetric spin-dependent electric dipole}

The expectation value of the spin dependent electric dipole of Co(1) at the bond-\textcircled{\scriptsize 1} [see Fig. \ref{fig:crystal}(a)]
can be expressed in the following general form
\cite{Kaplan-2011,Miyahara-2016}:
\begin{align}
\Bigl( p^x,~p^y,~p^z \Bigr)_{\textcircled{\scriptsize 1}}
=
\Bigl( a_1 W_x + a_2 W_y + a_3 W_z,~~b_1 W_x + b_2 W_y + b_3 W_z,~~c_1 W_x + c_2 W_y + c_3 W_z \Bigr).
\label{eqn:w1}
\end{align}
Here, $a_i$, $b_i$, and $c_i$ ($i=1,2,3$) are coefficients.
$\bW=\braket{\bS_{\rm A}}\times\braket{\bS_{\rm B}}$ is the expected value of the vector spin chirality,
and $W_\alpha$ represents its $\alpha(=x,y,z)$ component.
Notice that Eq. (\ref{eqn:w1}) provides the general form of the antisymmetric spin-dependent electric dipole,
since Co(1) only possesses the space inversion and there is no restriction between the coefficients.
There are three kinds of bonds [see Fig. \ref{fig:crystal}(a)].
As in Eq. (\ref{eqn:w1}), the electric dipole at the bond-\textcircled{\scriptsize 2} is described with the same coefficients as
\begin{align}
\Bigl( p^{x'},~p^{y'},~p^{z'} \Bigr)_{\textcircled{\scriptsize 2}}
=
\Bigl( a_1 W_{x'} + a_2 W_{y'} + a_3 W_{z'},~~b_1 W_{x'} + b_2 W_{y'} + b_3 W_{z'},~~c_1 W_{x'} + c_2 W_{y'} + c_3 W_{z'} \Bigr).
\label{eqn:w2}
\end{align}
Here, $x'$, $y'$, and $z'$ represent the components in the $x'y'z'$ coordinate
which is rotated by 120$^\circ$ around the $z$-axis from the global $xyz$ coordinate.
Between the two coordinates, there are following relations:
\begin{align}
\begin{pmatrix}
p^x \cr
p^y \cr
p^z
\end{pmatrix}_{\textcircled{\scriptsize 2}}
=
\begin{pmatrix}
\cos\phi  & -\sin\phi & 0 \cr
\sin\phi & \cos\phi & 0 \cr
0 & 0 & 1
\end{pmatrix}
\begin{pmatrix}
p^{x'} \cr
p^{y'} \cr
p^{z'}
\end{pmatrix}_{\textcircled{\scriptsize 2}},
~~~~~~
\begin{pmatrix}
W_{x'} \cr
W_{y'} \cr
W_{z'}
\end{pmatrix}
=
\begin{pmatrix}
\cos\phi & \sin\phi & 0 \cr
-\sin\phi & \cos\phi & 0 \cr
0 & 0 & 1
\end{pmatrix}
\begin{pmatrix}
W_x \cr
W_y \cr
W_z
\end{pmatrix}.
\label{eqn:bond-w}
\end{align}
Here, $\phi=2\pi/3$.
In the same way, we can obtain the electric dipole at the bond-{\textcircled{\scriptsize 3}} with $\phi=-2\pi/3$.
Summing up the electric dipoles at the three bonds
($\bP=\bp_{\textcircled{\scriptsize 1}}+\bp_{\textcircled{\scriptsize 2}}+\bp_{\textcircled{\scriptsize 3}}$)
and using Eqs. (\ref{eqn:S}), (\ref{eqn:w1}), (\ref{eqn:w2}), and (\ref{eqn:bond-w}),
we obtain the following form of the electric polarization per a unitcell:
\begin{align}
\begin{pmatrix}
P^x \cr
P^y \cr
P^z
\end{pmatrix}
=3
\begin{pmatrix}
-S_\parallel S_\perp \cos\varphi \bigl[ (a_1+b_2)\cos\theta + (a_2-b_1)\sin\theta \bigr] \cr
-S_\parallel S_\perp \cos\varphi \bigl[ (a_1+b_2)\sin\theta - (a_2-b_1)\cos\theta \bigr] \cr
S_\perp^2 c_3 \sin2\varphi
\end{pmatrix}.
\label{eqn:p-w}
\end{align}
This indicates that the polarization rotates in the $xy$-plane with the rotation of the field ($\theta$-rotation)
\cite{Miyahara-2016},
as shown in Fig. \ref{fig:rotation}(a).
Since it is proportional to $S_\parallel S_\perp$, a finite magnetic field in the $z$ direction is required for the $\theta$-rotation.
$P^z$ is independent on $\theta$ and can be finite when the canting angle $\varphi\neq 0$.

For Co(1)', the result is related to that for Co(1) owing to the twofold axis.
The polarization is then expressed by Eq. (\ref{eqn:p-w})
with the replacement of $(a_1,a_2,b_1,b_2,c_3)\rightarrow (-a_1,a_2,b_1,-b_2,-c_3)$.
For Co(2), the twofold axis restricts the possible coefficients in Eq. (\ref{eqn:w1}).
The polarization is then expressed by Eq. (\ref{eqn:p-w})
with the replacement of $(a_1,a_2,b_1,b_2,c_3)\rightarrow (0,\tilde{a}_2,\tilde{b}_1,0,0)$
\cite{Kaplan-2011}.
Here, new coefficients with tilde are introduced for Co(2).
Co(2)' is related to Co(2) by the space inversion.
Since the polarization is described with the antisymmetric spin dependence, the result is the same as Co(2).

We summarize the results in Table \ref{table:antisymmetric} for the various Co sites.
Every site can exhibit the $\theta$-rotation.
Both the Co(1) and Co(1)' sites carry both the $(P^x,P^y)\propto(\cos\theta,\sin\theta)$ and $(-\sin\theta,\cos\theta)$ components.
When the two results are added, only the latter component remains.
In contrast to this, there is only the latter component for Co(2) and Co(2)'.
Then, the total polarization only has the $(-\sin\theta,\cos\theta)$ component for the $\theta$-rotation.
Therefore, it is difficult to explain the observed $2\theta$-rotation within the antisymmetric spin-dependent electric dipole of the type-I theory.
Since the vector spin chirality $\bW=\braket{\bS_{\rm A}} \times \braket{\bS_{\rm B}}$ behaves as a vector for rotation,
it can describe the $\theta$-rotation but fails in explaining the $2\theta$-rotation.
For the $2\theta$-rotation, quadrupole degrees of freedom, such as $x^2-y^2$ or $xy$ type, are required.
In the following subsection, we investigate the symmetric spin-dependent electric dipole,
where such quadrupole degrees of freedom are present.

\begin{table}[t]
\begin{center}
\caption{
Electric polarization induced by the antisymmetric spin-dependent electric dipole (type-I theory) at various Co sites.
We introduced
$f_1=3S_\parallel S_\perp (a_1+b_2)$,
$f_2=3S_\parallel S_\perp (a_2-b_1)$, and
$\tilde{f}_2=3S_\parallel S_\perp (\tilde{a}_2-\tilde{b}_1)$.
Here, ($a_1$, $a_2$, $b_1$, $b_2$, $c_3$) and ($\tilde{a}_1$, $\tilde{a}_2$, $\tilde{b}_1$, $\tilde{b}_2$, $\tilde{c}_3$)
are coefficients for (Co(1), Co(1)') and (Co(2), Co(2)') sites, respectively.
The symbol ``$-$" represents that no polarization is induced.
As shown in Figs. \ref{fig:rotation} and \ref{fig:crystal}(a), $\theta$ represents the angle of the magnetic field,
whereas $\varphi$ is the canting angle of the magnetic structure [see Fig. \ref{fig:crystal}(b)].
}
\label{table:antisymmetric}
\begin{tabular}{lccc}
\hline
 & $P^x$ & $P^y$ & $P^z$ \cr
\hline 
Co(1) & $\cos\varphi(-f_1\cos\theta-f_2\sin\theta)$ & $\cos\varphi(-f_1\sin\theta+f_2\cos\theta)$ & $3S_\perp^2 c_3 \sin2\varphi$ \cr
Co(1)' & $\cos\varphi(f_1\cos\theta-f_2\sin\theta)$ & $\cos\varphi(f_1\sin\theta+f_2\cos\theta)$ & $-3S_\perp^2 c_3 \sin2\varphi$ \cr
Co(2) & $-\tilde{f}_2\cos\varphi\sin\theta$ & $\tilde{f}_2\cos\varphi\cos\theta$ & $-$ \cr
Co(2)' & $-\tilde{f}_2\cos\varphi\sin\theta$ & $\tilde{f}_2\cos\varphi\cos\theta$ & $-$ \cr
Total & $-2(f_2+\tilde{f}_2)\cos\varphi\sin\theta$ & $2(f_2+\tilde{f}_2)\cos\varphi\cos\theta$ & $-$ \cr
\hline
\end{tabular}
\end{center}
\end{table}

\subsection{Symmetric spin-dependent electric dipole}

In the absence of the inversion center, the symmetric spin dependence is allowed in the electric dipole.
First, we begin with the following general form of the electric dipole:
\cite{Matsumoto-2017}:
\begin{align}
\begin{pmatrix}
p^x \cr
p^y \cr
p^z
\end{pmatrix}_{\textcircled{\scriptsize 1}}
=
\begin{pmatrix}
A_1 F_{x^2} + A_2 F_{y^2} + A_3 F_{z^2} + A_4 F_{yz} + A_5 F_{zx} + A_6 F_{xy} \cr
B_1 F_{x^2} + B_2 F_{y^2} + B_3 F_{z^2} + B_4 F_{yz} + B_5 F_{zx} + B_6 F_{xy} \cr
C_1 F_{x^2} + C_2 F_{y^2} + C_3 F_{z^2} + C_4 F_{yz} + C_5 F_{zx} + C_6 F_{xy}
\end{pmatrix}.
\label{eqn:p1}
\end{align}
Here, $A_i$, $B_i$, and $C_i$ ($i=1-6$) are coefficients.
$F_{\alpha\beta}=\braket{S_\rA^\alpha} \braket{S_\rB^\beta} + \braket{S_\rA^\beta} \braket{S_\rB^\alpha}$ for $\alpha\neq\beta$,
and $F_{\alpha^2}=\braket{S_\rA^\alpha} \braket{S_\rB^\alpha}$ with Eq. (\ref{eqn:S}).
At the bond-\textcircled{\scriptsize 2}, the electric dipole is expressed in the rotated $x'y'z'$ coordinate, as in Eq. (\ref{eqn:w2}),
with the same coefficients in Eq. (\ref{eqn:p1}).
Between the two coordinates, the expectation value of the spin operator has the same relation as in Eq. (\ref{eqn:bond-w})
with the replacement of $(W^{\alpha'},W^\alpha)\rightarrow (\braket{S_i^{\alpha'}},\braket{S_i^\alpha})$ ($i={\rm A},{\rm B}$).
Summing up the electric dipoles at the three bonds as in the antisymmetric case,
we obtain the following form of the electric polarization per a unitcell:
\begin{align}
\begin{pmatrix}
P^x \cr
P^y \cr
P^z
\end{pmatrix}
=
\begin{pmatrix}
- g_1 \sin\varphi \sin\theta + g_1' \sin\varphi \cos\theta - g_2 \sin2\theta - g_2' \cos2\theta \cr
g_1 \sin\varphi \cos\theta + g_1' \sin\varphi \sin\theta - g_2 \cos2\theta + g_2' \sin2\theta \cr
3 C_3 S_\parallel^2 - \frac{3}{2}(C_1+C_2) S_\perp^2 \cos2\varphi
\end{pmatrix}.
\label{eqn:p-general}
\end{align}
Here,
$g_1=3S_\parallel S_\perp (A_5-B_4)$,
$g_1'=3S_\parallel S_\perp (A_4+B_5)$,
$g_2=\frac{3}{4}S_\perp^2 (A_1-A_2-2B_6)$,
and
$g_2'=\frac{3}{4}S_\perp^2 (2A_6+B_1-B_2)$.
The $(P^x,P^y)\propto (-\sin\theta,\cos\theta)$ and $(\cos\theta,\sin\theta)$ terms
represent the $\theta$-rotation [see Fig. \ref{fig:rotation}(a)].
It appears when the field has a finite $z$ component ($S_\parallel \neq 0$).
Since it is proportional to $\sin\varphi$, $(P^x,P^y)$ changes its sign when the field is reversed as $H\rightarrow -H$
[see Fig. \ref{fig:crystal}(b) for the field-sweeping process].
The $(P^x,P^y)\propto (\sin2\theta,\cos2\theta)$ and $(-\cos2\theta,\sin2\theta)$ terms in Eq. (\ref{eqn:p-general}) represent
that the polarization rotates in the opposite direction at the twice speed relative to the rotation of the external magnetic field,
i.e. $2\theta$-rotation as shown in Fig. \ref{fig:rotation}(b).
Since it is independent on the canting angle $\varphi$, the $(P^x,P^y)$ does not change under the field-sweeping process.
It is clear that the $\theta$-rotation and the $2\theta$-rotation
originate from the $(F_{yz},F_{zx})$ and $(F_{x^2},F_{y^2},F_{xy})$ spin dependences
in Eq. (\ref{eqn:p1}), respectively.

Next, we consider the Co(2) site with the twofold axis around the $y$ direction.
In this case, the twofold symmetry restricts the possible spin dependences,
and the following coefficients must vanish in Eq. (\ref{eqn:p1}): $A_4=A_6=B_1=B_2=B_3=B_5=C_1=C_2=C_3=C_5=0$
\cite{Matsumoto-2017}.
This indicates that $P^z=0$ and $g_1'=g_2'=0$ in Eq. (\ref{eqn:p-general}).
We can see that both the $\theta$-rotation and $2\theta$-rotation remain in $(P^x,P^y)$.
As shown in Fig. \ref{fig:crystal}(c), Co(2)' is related to Co(2) by the space inversion.
Since the polarization is described with the symmetric spin dependence,
the polarization for Co(2)' is expressed by Eq. (\ref{eqn:p-general}) with the replacement of
$(A_i,,B_i,C_i)\rightarrow (-A_i,-B_i,-C_i)$.
We summarize the results in Table \ref{table:symmetric} for the various Co sites.
For Co(1) and Co(1)', notice that no symmetric spin-dependent electric dipole is allowed owing to the inversion center.

\begin{table}[t]
\begin{center}
\caption{
Electric polarization induced by the symmetric spin-dependent electric dipole (type-I theory) at various Co sites.
``General" in the table represents the site with no symmetry transformations.
Notice that the Co(1) and Co(1)' sites possess the inversion center,
whereas the Co(2) and Co(2) sites possess the twofold axis around the $y$ direction.
We introduced
$g_1=3S_\parallel S_\perp (A_5-B_4)$,
$g_1'=3S_\parallel S_\perp (A_4+B_5)$,
$g_2=\frac{3}{4}S_\perp^2 (A_1-A_2-2B_6)$,
$g_2'=\frac{3}{4}S_\perp^2 (2A_6+B_1-B_2)$,
$g_3=3C_3 S_\parallel^2$,
$g_4=\frac{3}{2}(C_1+C_2) S_\perp^2$.
Here, $A_i$, $B_i$, $C_i$ ($i=1-6$) are independent coefficients introduced in Eq. (\ref{eqn:p1}) for the general description.
}
\label{table:symmetric}
\begin{tabular}{cccc}
\hline
 & $P^x$ & $P^y$ & $P^z$ \cr
\hline
General & $-g_1\sin\varphi\sin\theta + g_1'\sin\varphi\cos\theta$ & $g_1\sin\varphi\cos\theta+g_1\sin\varphi\sin\theta$ & $g_3-g_4\cos2\varphi$ \cr
 & $-g_2\sin2\theta-g_2'\cos2\theta$  & $-g_2\cos2\theta+g_2'\sin2\theta$ & \cr
\hline
Co(1) \& Co(1)' & $-$ & $-$ & $-$ \cr
Co(2) & $g_1\sin\varphi\sin\theta+g_2 \sin2\theta$ & $-g_1\sin\varphi\cos\theta+g_2 \cos2\theta$ & $-$ \cr
Co(2)' & $-g_1\sin\varphi\sin\theta-g_2 \sin2\theta$ & $g_1\sin\varphi\cos\theta-g_2 \cos2\theta$ & $-$ \cr
Total & $-$ & $-$ & $-$ \cr
\hline
\end{tabular}
\end{center}
\end{table}

\section{Summary and Discussions}

Let us discuss the reason why the polarization contains both the $\theta$-rotation and the $2\theta$-rotation.
This is owing to the fact that the physical quantities must be equivalent
when the magnetic field is rotated by 120$^\circ$ around the $z$-axis under the $C_3$ point group symmetry.
Thus, the electric polarization must rotate 120$^\circ$ when the field is rotated by 120$^\circ$.
The $\theta$-rotation satisfies this condition.
In case of the $2\theta$-rotation, interestingly, the electric polarization rotates $-240^\circ$,
and it is equivalent to $+120^\circ$ rotation.
Thus, the $2\theta$-rotation also satisfies the condition.
This is the reason why both the $\theta$-rotation and the $2\theta$-rotation are allowed in the electric polarization
under the $C_3$ point group symmetry
\cite{Matsumoto-2019}.

For the type-I theory, the antisymmetric spin-dependent electric dipole
does not possess the $2\theta$-rotation (see Table \ref{table:antisymmetric}),
whereas the symmetric one possesses it (see Table \ref{table:symmetric}).
The latter seems to explain the observed magnetoelectric effects,
however, the total polarization vanishes by the cancellation of the Co(2) and Co(2)' sites, owing to the inversion center between the two sites.
Therefore, it is difficult to explain the observed $2\theta$-rotation in \Co~within the type-I theory,
where the electric dipole is described by the product of spin operators at different sites.

The lines at ``Co(1)" and ``General" in Tables \ref{table:antisymmetric} and \ref{table:symmetric}, respectively,
give general description of the electric polarization on honeycomb lattices.
In a specific material, they can be used under consideration of symmetry transformations
at the bond-\textcircled{\scriptsize 1} [see Fig. \ref{fig:crystal}(a)].
This restricts the active spin-dependences in the polarization,
and leads to a reduction of several coefficients in Eqs. (\ref{eqn:w1}) and (\ref{eqn:p1}).
When $g_2\neq 0$ or $g_2'\neq 0$ in Table \ref{table:symmetric}, there remains the $2\theta$-rotation terms for the type-I theory.
These terms are cancelled out in \Co, however, it can remain when materials have no inversion center.

\begin{table}[t]
\begin{center}
\caption{
Electric polarization induced by product of spin operators at the same site (type-II theory)
\cite{Matsumoto-2019}.
$K_1$, $K_1'$, $K_2$, and $K_2'$ are independent coefficients.
$O_1$ is amplitude of the expectation value of $O^{yz}=S^y S^z+S^y S^z$ or $O^{zx}=S^z S^x+S^x S^z$ type quadrupoles,
whereas $O_2$ is for $O^{x^2-y^2}=(S^x)^2- (S^y)^2$ or $O^{xy}=S^x S^y+S^y S^x$ type.
}
\label{table:local}
\begin{tabular}{lccc}
\hline
 & $P^x$ & $P^y$ & $P^z$ \cr
 \hline 
Co(1) & $2O_1 \cos\varphi(K_1 \sin\theta - K_1' \cos\theta )$ & $2O_1 \cos\varphi(-K_1 \cos\theta - K_1' \sin\theta )$ & $-$ \cr
 & $+2O_2 \sin2\varphi(K_2 \sin2\theta - K_2' \cos2\theta )$ & $+2O_2 \sin2\varphi(K_2 \cos2\theta + K_2' \sin2\theta )$ & $-$ \cr
\hline
Co(1)' & $2O_1 \cos\varphi(K_1 \sin\theta + K_1' \cos\theta )$ & $2O_1 \cos\varphi(-K_1 \cos\theta + K_1' \sin\theta )$ & $-$ \cr
 & $+2O_2 \sin2\varphi(K_2 \sin2\theta + K_2' \cos2\theta )$ & $+2O_2 \sin2\varphi(K_2 \cos2\theta - K_2' \sin2\theta )$ & $-$ \cr
\hline
Co(2) & $2O_1 ( \tilde{K}_1 \cos\varphi + \tilde{K}_1' \sin\varphi ) \sin\theta$ & $-2O_1 ( \tilde{K}_1 \cos\varphi + \tilde{K}_1' \sin\varphi )\cos\theta$ & $-$ \cr
 & $+2O_2 ( \tilde{K}_2 \sin2\varphi - \tilde{K}_2' \cos2\varphi ) \sin2\theta$ & $+2O_2 ( \tilde{K}_2 \sin2\varphi - \tilde{K}_2' \cos2\varphi )\cos2\theta$ & $-$ \cr
\hline
Co(2)' & $2O_1 ( \tilde{K}_1 \cos\varphi - \tilde{K}_1' \sin\varphi ) \sin\theta$ & $-2O_1 ( \tilde{K}_1 \cos\varphi - \tilde{K}_1' \sin\varphi )\cos\theta$ & $-$ \cr
 & $+2O_2 ( \tilde{K}_2 \sin2\varphi + \tilde{K}_2' \cos2\varphi ) \sin2\theta$ & $+2O_2 ( \tilde{K}_2 \sin2\varphi + \tilde{K}_2' \cos2\varphi )\cos2\theta$ & $-$ \cr
\hline
Total & $4O_1 ( K_1 + \tilde{K}_1 ) \cos\varphi\sin\theta$ & $-4O_1 ( K_1 + \tilde{K}_1 ) \cos\varphi\cos\theta$ & $-$ \cr
 & $+4O_2 ( K_2 + \tilde{K}_2 ) \sin2\varphi\sin2\theta$ & $+4O_2 ( K_2 + \tilde{K}_2 ) \sin2\varphi\cos2\theta$ & $-$ \cr
\hline
\end{tabular}
\end{center}
\end{table}

When the magnetic ion occupies a site lacking the inversion symmetry as in honeycomb lattices,
the electric dipole can be described by the product of spin operators at the same site (type-II theory),
and we also present Table \ref{table:local} for the type-II theory
\cite{Matsumoto-2019}.
Notice that Tables \ref{table:antisymmetric}, \ref{table:symmetric}, and \ref{table:local} are general
and applicable to other honeycomb antiferromagnets such as
BaNi$_2$(PO$_4$)$_2$ \cite{BaNiAsO}, BaNi$_2$V$_2$O$_8$ \cite{BaNiVO}, and MnPS$_3$ \cite{MnPS}.
For \Co, we clarified that only the type-II theory accounts for both the $2\theta$-rotation and the field-sweeping process.
We conclude that they originate from the local quadrupoles of the $O^{x^2-y^2}=(S^x)^2-(S^y)^2$ and $O^{xy}=S^x S^y+S^y S^x$ types
\cite{Matsumoto-2019}.
This is because that
$O^{x^2-y^2}({\rm A}) - O^{x^2-y^2}({\rm B})$ and $O^{xy}({\rm A}) - O^{xy}({\rm B})$
combinations can be realized
\cite{Kaplan-2011}.
Here, $O^i({\rm A})$ and $O^i({\rm B})$ $(i=x^2-y^2,xy)$ represent the quadrupoles at the A and B sites,
respectively [see Figs. \ref{fig:crystal}(a) and \ref{fig:crystal}(c)].
Since they are antisymmetric with respect to the space inversion,
the electric dipole of the local quadrupole origin can survive even in the presence of the inversion center between the Co sites.
This is the reason why the $2\theta$-rotation remains in \Co~with the inversion center.
We finally summarize the possible field-induced magnetoelectric effects for general honeycomb lattices in Table \ref{table:summary}.

\begin{table}[t]
\begin{center}
\caption{
Possible field-induced magnetoelectric effects on honeycomb lattices.
For the antisymmetric spin dependence, we focus on the Co(1) site in Table \ref{table:antisymmetric} for the general description.
For the symmetric one, we focus on the ``General" site in Table \ref{table:symmetric}.
$H_\perp$ represents that the polarization is induced by the perpendicular component of the external magnetic field $(H_x,H_y)$.
$H_\parallel H_\perp$ represents that it is induced in the presence of both the parallel ($H_z$) and perpendicular $(H_x,H_y)$ components.
$H_\parallel$ or $H_\perp$ represents that it can be independently induced by one of those components.
The “Linear ME effect” represents that the polarization shows a linear dependence on the magnetic field $\bH$.
It appears in $P^z$, $\theta$-rotation, and $2\theta$-rotation components of the polarization.
Notice that they are proportional to $\sin\varphi$ or $\sin2\varphi$
in Tables \ref{table:antisymmetric}, \ref{table:symmetric}, and \ref{table:local}.
}
\label{table:summary}
\begin{tabular}{lcccc}
\hline
 & $\theta$-rotation & $2\theta$-rotation & $P^z$ & Linear ME effect \cr
\hline 
Type-I (Antisymmetric spin-dependence) & $H_\parallel  H_\perp$ & $-$ & $H_\perp$ & $P^z$ \cr
Type-I (Symmetric spin-dependence) & $H_\parallel H_\perp$ & $H_\perp$ & $H_\parallel$ or $H_\perp$ & $\theta$-rotation \cr
Type-II & $H_\parallel H_\perp$ & $H_\perp$ & $H_\parallel$ & $\theta$-rotation, $2\theta$-rotation \cr
\hline 
\end{tabular}
\end{center}
\end{table}





\begin{thebibliography}{9}

\bibitem{Fischer-1972}
E. Fischer, G. Gorodetsky, and R. M. Hornreich,
Solid State Commun. {\bf 10}, 1127 (1972).

\bibitem{Khanh-2016}
N. D. Khanh et al.,
Phys. Rev. B {\bf 93}, 075117 (2016).

\bibitem{Deng-2018}
G. Deng et al.,
Phys. Rev. B {\bf 97}, 085154 (2018).

\bibitem{Khanh-2017}
N. D. Khanh, N. Abe, S. Kimura, Y. Tokunaga, and T. Arima,
Phys. Rev. B {\bf 96}, 094434 (2017).

\bibitem{Solovyev-2016}
I. V. Solovyev and T. V. Kolodiazhnyi,
Phys. Rev. B {\bf 94}, 094427 (2016).

\bibitem{Yanagi-2018-1}
Y. Yanagi, S. Hayami, and H. Kusunose,
Physica B {\bf 536}, 107 (2018).

\bibitem{Yanagi-2018-2}
Y. Yanagi, S. Hayami, and H. Kusunose,
Phys. Rev. B {\bf 97}, 020404(R) (2018).

\bibitem{Matsumoto-2019}
\label{ref:Matsumoto-2019}
M. Matsumoto and M. Koga,
J. Phys. Soc. Jpn. {\bf 88}, 094704 (2019).

\bibitem{Moriya-1968}
T. Moriya,
J. Appl. Phys. {\bf 39}, 1042 (1968).

\bibitem{Katsura-2005}
H. Katsura, N. Nagaosa, and A. V. Balatsky,
Phys. Rev. Lett. {\bf 95}, 057205 (2005).

\bibitem{Arima-2007}
T. Arima,
J. Phys. Soc. Jpn. {\bf 76}, 073702 (2007).

\bibitem{Kaplan-2011}
T. A. Kaplan and S. D. Mahanti,
Phys. Rev. B {\bf 83}, 174432 (2011).

\bibitem{Matsumoto-2017}
M. Matsumoto, K. Chimata, and M. Koga,
J. Phys. Soc. Jpn. {\bf 86}, 034704 (2017).

\bibitem{Mims-1976}
W. B. Mims,
{\it The Linear Electric Field Effect in Paramagnetic Resonance} (Oxford University Press, U.K., 1976).

\bibitem{Miyahara-2016}
S. Miyahara and N. Furukawa,
Phys. Rev. B {\bf 93}, 014445 (2016).



\bibitem{BaNiAsO}
L. P. Regnault et al.,
J. Magn. Magn. Mater. {\bf 15-18}, 1021 (1980).

\bibitem{BaNiVO}
N. Rogado, Q. Huang, J. W. Lynn, A. P. Ramirez, D. Huse, and R. J. Cava,
Phys. Rev. B {\bf 65}, 144443 (2002).

\bibitem{MnPS}
E. Ressouche et al.,
Phys. Rev. B {\bf 82}, 100408(R) (2010).



\end{thebibliography}
\end{document}